# The Radio Galaxy 3C265 Contains a Hidden Quasar Nucleus[1]

Arjun Dey

Astronomy Dept., University of California at Berkeley, CA 94720, &

Institute of Geophysics & Planetary Physics, LLNL, Livermore, CA 94550

Hyron Spinrad

Astronomy Dept., University of California at Berkeley, CA 94720

dey/spinrad@astro.berkeley.edu

astro-ph/9512077   12 Dec 95

# ABSTRACT

We report the discovery of broad MgII emission from the high redshift radio galaxy 3C265 (z=0.81). We detect the broad line in the nuclear spectrum *and in the spatially extended galaxian component*, both near the nucleus and in the spectrum of an off-nuclear knot located 31 kpc south east of the nucleus of the galaxy. These data provide strong support for the simplest form of the unification hypothesis, that radio galaxies are quasars whose optical radiation is directed in the plane of the sky rather than into our line of sight. These data also strongly support the scattering model for the alignment of the UV continuum emission with the radio axis. In 3C265, if the axis of the anisotropically emitted UV continuum radiation is identified with the major axis of the radio source, then the observed rest frame UV continuum emission implies that the opening angle of the radiation cone is large (half angle $\approx 45°$).

We also derive a mass estimate of $8 \times 10^{10}$ M$_\odot$ for the central region of 3C265 from its rotation curve. The implied mass-to-light ratio is low (M/L $\sim 2$), and suggests that a significant fraction of the rest frame UV continuum emission from this galaxy is dominated by reprocessed radiation from the buried AGN. Finally, we detect the CaII$\lambda$3933 K line in absorption in the integrated spectrum of 3C265. This provides direct spectroscopic evidence for the existence of stars in a high redshift radio galaxy.

*Subject headings:* galaxies: active — galaxies: individual (3C265) – galaxies: quasars: general — scattering — radio continuum: galaxies

## 1. Introduction

The large range in morphological and spectroscopic properties observed in luminous active galactic nuclei (AGN) has led to the definition of a complicated and perplexing taxonomy. In the crudest sense, AGN can be divided into two classes based on whether they are radio loud or radio quiet. Each of these classes can be further subdivided based on whether the optical spectra of the AGN exhibit broad permitted lines (Type I spectra) or narrow permitted lines (Type II spectra). It has been suggested that this latter spectral subdivision is due not to a variation in the intrinsic properties of the AGN, but instead to the combined effects of obscuration and the relative orientation of the observer's line of sight to the axis of the anisotropic emission from the nuclear source (cf. Antonucci 1993). These 'unification theories' have attempted to unify Seyfert I's with Seyfert II's, and narrow line FR-II radio galaxies with steep spectrum radio loud quasars.

Of the vast array of AGN properties that are elegantly explained by the unification hypothesis, two deserve particular mention. The first is the fundamental observation that spectropolarimetry of the nuclear region of several Type II objects (from both the radio loud and radio quiet classes) show broad permitted lines (*i.e.*, a Type I spectrum) in their polarized light (Antonucci & Miller 1985, Miller & Goodrich 1990). In the context of the unification hypothesis, this observation can be naturally attributed to the presence of a hidden quasar nucleus in Type II objects which is obscured from our direct view, but rendered visible by "mirrors" of scattering clouds of dust and electrons.

The second observation that may pertain to the unification of radio galaxies and quasars is that the restframe UV morphologies of high redshift ($z \gtrsim 1$) radio galaxies tend to be aligned with their radio axes (McCarthy *et al.* 1987, Chambers, Miley & van Breugel 1987). Understanding the origin of this alignment effect is extremely important since radio galaxies are the most often used probes of galaxy evolution and cosmology (Lilly & Longair 1984, Spinrad 1986). The recent observations of polarized continuum emission, and the possible detection of polarized broad line emission from these distant aligned galaxies provide strong support for the hypothesis that the aligned UV continua are dominated by scattered light from a hidden quasar (Tadhunter *et al.* 1988, di Serego Alighieri *et al.* 1989, 1993, 1994; Cimatti *et al.* 1993).

In this paper, we present our observations of 3C265, a $V \approx 20$ powerful narrow emission line radio galaxy at z=0.81 (Kristian, Sandage & Katem 1978, Smith *et al.* 1979, Saslaw, Tyson & Crane 1978). The major axis of the restframe UV morphology of this galaxy is offset by roughly $45°$ from the major axis of its radio emission ($PA_{optical} = 147°$, $PA_{radio}^{E.Lobe} = 112°$, $PA_{radio}^{W.Lobe} = 103°$). Although this galaxy is not a particularly good example of the alignment effect, it was selected as a target because its integrated continuum emission is known to be polarized at $\approx 10\%$ at an angle perpendicular to the optical major axis of the galaxy (Jannuzi & Elston 1991, Jannuzi 1994). We have obtained a spectrum of the extended UV component and discovered broad MgII emission, a feature commonly observed in the spectra of quasars, from both the nuclear region of the galaxy *and from its extended component*. This observation provides one of the most direct pieces of evidence that radio galaxies have a quasar nucleus that is generally not observed because it is directed away from our line of sight, and possibly heavily obscured.

Throughout this paper we assume that $H_\circ = 50\,\mathrm{km\,s^{-1}\,Mpc^{-1}}$, $q_\circ = 0$. The scale at z=0.81 (the redshift of 3C265) is then 10.1 kpc/$''$. In comparison, for $H_\circ = 75\,\mathrm{km\,s^{-1}\,Mpc^{-1}}$, $q_\circ = 0.5$ the angular scale at z=0.81 is 5.5 kpc/$''$.

## 2. Observations

3C265 was observed using the Low Resolution Imaging Spectrometer (LRIS; Oke *et al.* 1995) at the cassegrain focus of the 10-m W. M. Keck Telescope on U.T. 14 March 1994. We used a 300 line/mm grating (blazed at 5000Å) and a 1$''$ width slit which resulted in an effective resolution FWHM of 10Å. The LRIS detector is a Tek $2048^2$ CCD with $24\mu m$ pixels, corresponding to a scale of $0''.214$ pix$^{-1}$. We obtained three spectra (with exposure times of 1500$s$, 1800$s$ and 1500$s$) in PA=150°; the slit PA was chosen to be parallel to the major axis of the galaxy (Fig. 1). The parallactic angle during our observations was $\approx 113°$, but since the observations were made at small airmass, the relative spectrophotometry should be fairly accurate. The seeing during the observations was $0''.7 - 0''.8$.

The data were corrected for overscan bias and flat–fielded using internal lamps taken immediately following the observations. Flux calibration was performed using observations of HZ 21 (Oke 1974). The star was observed both with and without an order sorting OG570 filter in order to correct for the second order light contamination in the spectral region $\lambda > 7500$Å. Since there were problems with the CCD response and red side calibration on this Keck LRIS observing run, we also obtained spectroscopic observations of 3C265 with the Kast cassegrain double spectrograph at the 3-m Shane telescope of Lick Observatory. We obtained two 1800$s$ exposures with a wide (3$''$) slit in order to check the flux calibration of the Keck data. All the spectroscopic reductions were performed using the NOAO IRAF package.

## 3.   Results

The scope of the present work is primarily restricted to the spectral region around the MgII$\lambda$2799 line, which is shown in Fig. 2. The complex of lines near $\lambda \approx 2800$Å consists of at least 6 narrow lines: HeII$\lambda$2733, [MgVII]$\lambda$2786, MgII$\lambda$2796,2805, OIII$\lambda$2826,2835. The OIII lines are due to Bowen fluorescence with HeII, and will be discussed elsewhere. The typical width of narrow emission lines in 3C265 is $\approx 800 \, \mathrm{km \, s^{-1}}$. Note that the narrow line complex appears to be superimposed on a broad hump; this is the broad component to the MgII$\lambda$2799 line, which can be seen in the nuclear data even without subtraction of the narrow line components!

In order to study the spatially extended component of the galaxy, we extracted several one-dimensional spectra from our two dimensional image. All of these extractions were constrained to have the same linear model fit to the S-distortion of the two-dimensional image, and only differed in position along the slit and extraction width. We chose to adopt this method of analysing the two dimensional spectrum rather than studying the two dimensional image directly because the signal to noise ratio is low for the individual rows, and the individual rows are not independent because of seeing and any possible tracking errors (the scale is $0\rlap{.}''214$/pixel and the seeing was $0\rlap{.}''8 - 1\rlap{.}''0$). We therefore extracted 12 individual spectra from the two dimensional image; the aperture centers and extraction windows are listed in Table 1.

In order to determine the parameters of the broad component of the line we have fit each of the narrow lines by a gaussian model. The narrow lines were all constrained to have the same gaussian width, and the relative locations of the peaks were fixed by their vacuum wavelength ratios. Hence the free parameters are the total intensities of the narrow lines and the width of a single narrow line. Not all the one-dimensional extractions show all 6 spectral lines mentioned above in the spectral region around the MgII lines. As a result, only the strongest features were fit for each of the extractions. In particular, for the spectrum of the off-nuclear south-east knot only two narrow lines (those of MgII) were fit and subtracted. Figures 2, 3 and 4 show three representative examples of the spectral fits. In these figures, the upper panels show the total spectrum and the fit, and the lower panels show the broad component after subtraction of the narrow emission lines. The spectral analyses were performed using the SPECFIT software in IRAF (Kriss 1994). The derived parameters of the broad MgII emission line (flux, FWHM, equivalent width) and the formal errors are presented in Table 1.

## 4. Discussion

### 4.1. The Broad MgII Emission Line

Figures 5, 6 and 7 show the spatial variation along the slit of the FWHM, flux and equivalent width of the broad component of the MgII emission line. The horizontal bars show the width of the spectral extractions. Note that the FWHM of the broad MgII line is approximately constant for the three nuclear extractions with different widths. Moreover, the flux in the broad component increases with increasing extraction width, suggesting that the broad MgII emission is spatially extended. This is confirmed by the off-nuclear spectral extractions centered $0\rlap{.}''9$ away from the nucleus which show broad line fluxes that are $\approx 50\%$ of the nuclear flux, almost four times more than that expected from the effects of seeing. The equivalent width of the broad MgII component also remains roughly constant for the spectral extractions centered on and near the nucleus; this implies that the broad component has a spatially extended component near the nucleus, and, if it is scattered light, that the underlying continuum is also scattered in a similar fashion.

Finally, it is remarkable that not only the extended component near the nucleus, but an off-nuclear knot, located $3\rlap{.}''1$ (31 kpc) south east of the nucleus *also shows evidence for a broad component to the MgII line, with a FWHM very similar to that of the nuclear broad component.* There is only one process by which an off-nuclear region could exhibit a broad MgII line, and that is by scattering the nuclear AGN spectrum off dust or electrons. The off nuclear knot has a broad component of the MgII line with a flux which is approximately three times less than the flux observed from the nucleus, and an equivalent width which is a factor of two less than that in the nucleus. This suggests that both the continuum and the broad emission line are varying in a comparable manner. This is expected if most of the extended continuum emission associated with the galaxy and the off-nuclear knot is also scattered AGN light.

If the observed broad MgII line and continuum emission associated with the off-nuclear knot are indeed scattered light from a hidden quasar nucleus, the equivalent width of the broad line, the measured luminosity in the line and the specific luminosity of the continuum should be comparable to that observed in quasars at similar redshifts. The rest frame equivalent width of the broad MgII line in the off-nuclear knot is $\approx 24$Å, which is well within the range observed in radio loud quasars in the redshift range $0.7 \lesssim z \lesssim 1.5$ (15Å $\lesssim W_{rest}^{qso} \lesssim 60$Å; Steidel and Sargent 1991).

The luminosity in the broad MgII line from the knot is observed to be $\approx 3.4 \times 10^{40}$ergs s$^{-1}$ sterad$^{-1}$. If the broad MgII observed in the spectrum of the knot is scattered light from the nucleus, then the luminosity of the broad line component from the nucleus is given by

$$L_{\text{broad MgII}}^{nuc} = L_{\text{broad MgII}}^{knot} \left( \frac{D_{proj}}{R \cos\theta} \right)^2 \epsilon^{-1}$$

where $D_{proj}$ is the projected linear distance between the knot and the nucleus (3$''$1 or 31.5 kpc), $R$ is the size of the scattering cloud (1$''$ or 10 kpc), $\theta$ is the angle between the radius vector from the nucleus to the cloud and the plane of the sky, and $\epsilon$ is the scattering probability, which depends upon the type, number density and filling factor of the scattering particles. If the scattering is optically thin, $\epsilon \approx \tau_s f_C g(\frac{\pi}{2} - \theta)$ where $\tau_s$ is the scattering optical depth, $f_C$ is the covering fraction, and $g(\theta)$ is the scattering phase function, the probability that a photon will be scattered by an angle $\theta$ into our line of sight.

The size of the scattering knot can be estimated from the recent HST observations of Longair, Best & Röttgering (1995) and is found to be $\approx 1''$ or $\approx 10$ kpc. The off-nuclear knot is located $\approx 3''$1 (or $\approx 31$ kpc) from the nucleus, and therefore (for 100% scattering efficiency) the intrinsic luminosity of the nuclear broad MgII emission line must be at the very minimum $L_{\text{broad MgII}}^{nuc} > 3.3 \times 10^{41}$ ergs s$^{-1}$ sterad$^{-1}$.

It is important to note that this estimate lies well below the range of luminosities observed in quasars. Typical luminous radio loud quasars in the redshift range $0.7 \lesssim z \lesssim 1.5$ have broad MgII luminosities in the range $L_{\text{broad MgII}}^{nuc} \sim 10^{43.2-44.3}$ ergs s$^{-1}$ (Steidel & Sargent 1991; Osterbrock 1989; Brotherton et al. 1994). If the nuclear source has a luminosity of $10^{44}$ ergs s$^{-1}$ sterad$^{-1}$, the implied scattering efficiency is $\epsilon \approx 0.3\%$.

The two most likely scattering processes are Thomson scattering by electrons and dust scattering. For Thomson scattering, the optical depth is $\tau_{es} = \langle n_{es} R \rangle \sigma_T = 0.002 \langle n_{es} R_{\text{kpc}} \rangle$ where $\sigma_T$ is the Thomson cross section, $n_{es}$ is the electron density in units of cm$^{-3}$, $R_{\text{kpc}}$ is the size of the scattering cloud in kpc. Assuming that the size of the off-nuclear knot is $\approx 10$ kpc and an electron density of $n_e \approx 1$ cm$^{-3}$, the luminosity of the broad MgII emission line from the nucleus is $L_{\text{broad MgII}}^{nuc} \approx 2.8 \times 10^{43}$ erg s$^{-1}$ sterad$^{-1}$. Here we have assumed that $f_C \approx 1$ which is undoubtedly a gross overestimate. In fact, the HST image of 3C265 clearly shows that the off–nuclear knot shows substructure and is comprised of at least two sub–clumps. Nevertheless, the assumption of $f_C \approx 1$ provides us with some idea of the implied minimum luminosity for the central source.

If electrons are indeed responsible for scattering the broad MgII line, the width of the line places a constraint on the electron temperature. The observed FWHM of the broad MgII line is roughly the same both in the extended nuclear component and in the off–nuclear knot (FWHM$_{\text{MgII}} \approx 9000$ km s$^{-1}$). Moreover, the FWHM of the observed broad component both in the nuclear and off–nuclear components is within the range of FWHM of the broad MgII line commonly observed in quasars. In radio loud quasars the distribution of $FWHM_{MgII}$ is found to be broad, ranging between 2100 km s$^{-1}$ and 10300 km s$^{-1}$ with a mean value of $\langle FWHM_{MgII,QSO} \rangle \approx 4620 \pm 310$ km s$^{-1}$ (Brotherton *et al.* 1994). If we assume that the nuclear source in 3C265 has an intrinsic $FWHM_{MgII} \geq 2100$ km s$^{-1}$ and that the observed emission line from the extended component of the galaxy is broadened due to scattering by a hot thermal population of electrons, then the temperature of the electrons is constrained to be

$$T_e \leq 3.965 \times 10^{-3} \left( FWHM^2_{MgII,3c265} - FWHM^2_{MgII,QSO} \right)$$

or, in the case of 3C265, $T_e \leq 3 \times 10^5$ K. The fact that the FWHM of the broad component does not increase with radius places an even tighter constraint on the presence of a hot electron halo. This implies that the line is not additionally broadened due to scattering by a hot ($T_e \gtrsim 10^7$ K) population of thermal electrons, as has been suggested by Fabian (1989) for cooling flow electrons deposited upon large galaxies. Hence, if electrons are indeed responsible for scattering the nuclear light, the population must be cooler than $3 \times 10^5$ K. It is intriguing that this upper limit is comparable to the temperature of the electron scattering population in the nuclear regions of the nearby Seyfert galaxy NGC 1068 (Miller, Goodrich & Matthews 1991).

Similar arguments may be used to estimate the luminosity of the intrinsic nuclear continuum emission. It is likely that the continuum is also dominated by scattered light, since the equivalent width of the broad MgII line is similar to that observed in luminous quasars. The observed continuum emission at $\lambda 5500$Å from the off-nuclear knot is $\approx 2 \times 10^{-29}$ erg s$^{-1}$ cm$^{-2}$ Hz$^{-1}$ (or $V \approx 23$). Therefore the continuum luminosity of the knot is $\sim \nu f_{cont} d_L^2 \approx 4.8 \times 10^{42}$ erg s$^{-1}$ sterad$^{-1}$. Assuming that the scattering is 100% efficient provides a lower limit to the incident nuclear luminosity of $L^{nuc}_{continuum} \gtrsim 4.6 \times 10^{43}$ erg s$^{-1}$ sterad$^{-1}$. This is well below the range observed for $z \approx 1$ quasars, which have continuum luminosities at $\lambda_{rest} \approx 3000$Å of $\nu L_\nu (3000$Å$) \sim 10^{45-46.5}$ erg s$^{-1}$ sterad$^{-1}$ (Steidel and Sargent 1991, Osterbrock 1989).

Another constraint on the scattering mechanism can be derived from the polarization observations of Jannuzi and Elston (1991). If we assume that the primary scattering mechanism is Thomson scattering within a radiation cone of half opening angle 45° uniformly filled with electrons, then the maximum possible intrinsic polarization of the scattered light is $P_{max} \approx 53\%$. Since the observed percentage polarization of the spatially extended continuum emission is roughly 30% in the $B$-band (Jannuzi and Elston 1991), we estimate the luminosity of the scattered component to be at least $5.5 \times 10^{44}$ erg s$^{-1}$. If we assume that the nucleus is a very luminous $10^{46}$ erg s$^{-1}$ quasar, then the implied scattering efficiency requires that the average electron density in the extended component (within 30 kpc) be $\approx 3.7$ cm$^{-3}$. This results in an unrealistically large estimate of $\approx 10^{13}$ M$_\odot$ for the mass of the *ionized* component in the galaxy (within 30 kpc), and it is therefore unlikely that the dominant scattering mechanism is electron scattering. Similar arguments have been used to rule out electron scattering in other $z \approx 1$ radio galaxies (di Serego Alighieri *et al.* 1994, Dey *et al.* 1995).

The present data are unable to conclusively distinguish between dust and electron scattering. However, the efficiency of scattering implied by the continuum luminosity of the off–nuclear knot suggest that dust particles are far more likely a candidate than electrons. Dust scattering tends to be more efficient than electron scattering for two reasons. First, and most important, the scattering cross section for dust in the UV and optical spectral regime is of the same order as the geometric cross section of the dust particles. Second, if the optical and near–UV light ($\lambda_\circ > 3000$Å) is Rayleigh scattered, the scattering is more efficient in the UV. The crucial discriminatory test will be spectropolarimetry of the continuum spectrum of the spatially extended component of the galaxy and the extra–nuclear knot. In addition, near infrared spectroscopy will permit us to search for a broad component to the Balmer line emission and thereby place limits on the reddening in the nuclear broad line region and perhaps the scattering process. These data will allow us to study the spectrum of the hidden quasar and possibly probe the properties of dust particles at a redshift of 0.8.

In summary, the observed properties (FWHM, equivalent width, luminosities) of the nuclear *and off–nuclear* MgII broad line emission from 3C265 are consistent with the properties of broad line emission observed in quasars. This observation, in combination with the recent detection of polarization associated with the MgII line in the integrated spectra of two radio galaxies (di Serego Alighieri *et al.* 1994), provides the strongest evidence to date that at least some radio galaxies harbour "mis-directed" quasar nuclei as envisioned by the AGN unification hypothesis.

This observation also provides an important piece of evidence for our understanding of the alignment effect. Since some radio galaxies can contain buried, mis–directed quasars, the extended optical continuum emission of radio galaxies may contain a significant fraction of scattered light. The scattering of nuclear AGN light would therefore provide a natural explanation for the alignment effect. The one fly in the ointment is that 3C265 is *not* a good example of the alignment effect: the optical morphology of the galaxy is oriented almost 45° from the major axis of the radio emission. This has two possible implications: either the optical/UV radiation axis has little relation to the radio axis (which presumably defines the axis of bulk outflow from the AGN), or the opening angle of the optical/UV radiation cone is large (half − angle > 45°). In the few (8) known nearby Seyfert galaxies with ionization cones, the axis of the cones tend to be very well aligned with the axis of radio emission (Wilson & Tsvetanov 1994, and references therein). However, the opening angles of the ionization cones are large ($\approx 50°$ to $90°$). If 3C265 is similar to the low redshift radio Seyferts, then it is most likely that the opening angle of the nuclear radiation cone is large (opening angle $\approx 90°$). This hypothesis is testable: if the cone angle is large and the axis of the cone coincides with the radio axis, then the companion objects located to the east of 3C265 may lie within the radiation cone and their spectra should show a weak scattered broad MgII component.

## 4.2. The Velocity Structure of Ionized Gas in 3C265

The availability of longslit spectroscopic data of such unprecedented quality on a high redshift galaxy permit detailed studies of the kinematics of the gas associated with this galaxy. The velocity structure of the [OII]$\lambda$3727 emission line in 3C265 is seen in figure 8 which shows the two-dimensional spectrum of the galaxy in the vicinity of the emission line. The abscissa represents the distance along the slit in PA=150°; the origin is chosen to be the location of the peak of the continuum emission from the galaxy. A linear model was fit to the galaxy continuum in the vicinity of the [OII] line and subtracted so as not to bias the analysis of the velocity profile. The off–nuclear knot from which the broad MgII emission line is scattered also shows [OII] emission at roughly the systemic velocity of the galaxy.

The velocity profile derived from the [OII]$\lambda$3727 line is plotted in Fig. 9. This curve was determined by measuring the centroid of each CCD column of the [OII] line. Since the columns are spaced $0\farcs214$ apart, and the seeing during the observation was $\approx 0\farcs8$ at best, the individual points are not independent measures of the rotation curve but are useful in outlining the trends in the velocity profile. The errors in the measured centroids are reflected by the scatter in the points.

It is remarkable that this profile, unlike the velocity structure in most high redshift radio galaxies, is ordered, symmetric about the nucleus of the galaxy, and restricted to fairly small ranges in velocity ($\Delta v < 300$ kms$^{-1}$). These properties suggest that we are observing rotation in 3C265. The central portion of the galaxy ($|\theta| \leq 1\farcs5$, $\leq 15$ kpc) appears to rotate as a solid body. The 'rotation curve' then flattens for $4'' > |\theta| > 1\farcs5$ and then continues to rise at larger distances from the nucleus. The kinematic properties of the central region of the galaxy therefore appear to be 'disjoint' from the outer parts. The velocity profile of 3C265 is similar to that of some low redshift radio galaxies (*e.g.*, PKS 0634$-$206, 3C33, 3C227) that are classified as 'rotators' (Baum, Heckman & van Breugel 1992, 1990; Tadhunter, Fosbury & Quinn 1988). In this respect 3C265 may be a merger; indeed, the rotation curve of the galaxy is similar to the rotational profiles of low redshift galaxies thought to be merger remnants (*e.g.*, Schweizer 1990).

Do the kinematics of the gas reflect the gravitational potential? The presence of a powerful source of energy in the nucleus of a radio galaxy, and the existence of a large scale radio source which directly attests to some form of bulk outflow from the galaxy imply that it may be incorrect to interpret the velocity profile derived from the emission line gas as a rotation curve. However, our spectrum of 3C265 also shows a weak CaII$\lambda$3933 stellar absorption feature (figure 10), and although the signal to noise ratio on the red side of our spectrum is low, there is marginal evidence that the stars also participate in the rotation in the central regions of the galaxy. The centroid of the CaII K line in the SE region of the galaxy (*i.e.*, in a spectral extraction of width 0''.86 centered 0''.43 SE of the nucleus) is $\approx 8.8\text{Å}$ or $\approx 370\,\mathrm{km\,s^{-1}}$ different from the centroid of the absorption line in the NW region of the galaxy (*i.e.*, in a spectral extraction of width 0''.86 centered 0''.43 NW of the nucleus). This is in rough agreement with the central part of the [OII] velocity profile, and therefore strongly suggests that the emission line gas traces the galaxian rotation, at least in the central regions of the galaxy.

If we assume that the kinematics of the gas in the central regions of the galaxy are primarily influenced by gravity, then the rotation curve of the galaxy can provide a rough estimate of the galaxy's mass. If we restrict ourselves to the region in solid body rotation ($|\theta| \leq 1''.5$ or 15 kpc), and note that the galaxy is rotating at $\approx 150$ km s$^{-1}$ at a distance of $\approx 15$ kpc from the nucleus, the mass within this radius is $M\sin^3 i \sim \frac{v^2 R}{G} \approx 8 \times 10^{10}$ M$_\odot$ which is comparable to the mass of a large galaxy and therefore appears to be a reasonable estimate. If we calculate the mass using the outermost points on the 'rotation curve', this estimate increases by at most a factor of 10. This result is somewhat modulated by the fact that the mass of a rotating oblate spheroid is less than that estimated from the formula above. However, the correction due to geometry is no more than a factor of 2 (*e.g.*, Burbidge & Burbidge 1975), and is out-weighed by the uncertainties introduced by inclination effects and our lack of understanding of the detailed kinematics of the emission line gas and its relationship to the radio source.

By combining this estimate of the mass within 20 kpc with the nuclear luminosity of 3C265 we can estimate the mass-to-light ratio in this high redshift system. The flux density in the continuum at $\lambda \sim 6500\text{Å}$ is $\approx 2 \times 10^{-28}$ erg s$^{-1}$ cm$^{-2}$ Hz$^{-1}$. For H$_\circ$=50, q$_\circ$=0, this implies that the restframe blue luminosity of the galaxy is $\sim 4 \times 10^{10}$ L$_\odot$, implying a mass–to–light ratio of 2. This value is very low compared to most nearby early–type galaxies, and can only be attributed to the fact that the UV flux of the galaxy is dominated by light from the AGN; the stellar component, if present at UV wavelengths, must be weak.

This is confirmed by an examination of the spectrum of the galaxy which shows no stellar absorption lines in the continuum light except for the weak CaII K absorption feature (figure 10). The rest frame equivalent width of the CaII K absorption line is measured to be 5.5±0.5Å and the (rest frame) FWHM is 20Å. Note that the equivalent width measured here is a lower limit since a significant fraction of the continuum emission is probably reprocessed nonstellar emission from the AGN, and in addition the absorption feature may be contaminated by line emission. The width and FWHM are too high to be interstellar, and imply that the absorption line arises in stellar photospheres. The measured widths are similar to widths determined in the integrated spectra of nearby galaxies: for example, M31 has a CaII K line equivalent width of approximately 13Å and a FWHM of 21.5Å. Nearby E and S0 galaxies have distribution of CaII K equivalent width ranging from about 10Å to 20Å peaking at ≈17Å, whereas nearby spiral galaxies exhibit a very broad distribution ranging from ≈2Å to ≈20Å (*e.g.*, Alloin, Arimoto & Bica 1989). The somewhat lower equivalent width measured in the case of 3C265 than most early type galaxies may indicate dilution of the absorption line by the reprocessed and scattered nuclear continuum emission and *in situ* line emission.

## 5.  Conclusions

We have presented new long slit spectroscopic observations of the high redshift radio galaxy 3C265 (z=0.811) in the spectral region near the MgII$\lambda$2799 emission line. The spectra show clear evidence of broad line MgII emission in the nucleus of the galaxy *and in the spatially extended component*. We also detect broad MgII line emission in an off–nuclear knot located 31 kpc from the nucleus. The luminosity, FWHM and equivalent width of the broad line emission is typical of that observed in quasars. This observation, together with the detections of polarization from this and other high redshift radio galaxies (di Serego Alighieri *et al.* 1989, 1994), provides the clearest evidence to date that at least some radio galaxies harbour a quasar nucleus that is directed away from our line of sight, and support the unification hypothesis that radio galaxies are quasars with their anisotropic nuclear radiation directed away from our line of sight. The existence of scattered quasar light in the off–nuclear extended component of this radio galaxy also supports the hypothesis that the alignment effect observed in high redshift radio galaxies may be due to scattering of anisotropic nuclear continuum emission from a hidden AGN. Since the optical extent of 3C265 is not aligned with its radio axis, we are drawn to the conclusion that either the axis of nuclear optical emission can (on occasion) be strongly mis–aligned with the radio axis, or the opening angle of the radiation cone from the nucleus is large ($\gtrsim 45°$).

The discovery of hidden quasar nuclei in radio galaxies also bears on our ultimate understanding of the galaxian component of quasars. Ever since the discovery of quasars by Schmidt (1963), their galaxian properties have remained a mystery. Their prodigious luminosities that are produced in small ($\lesssim$ 1 kpc) regions generally overwhelm the light from any underlying galaxy. As a result, the host galaxies of powerful quasars have remained elusive, the only definitive spectroscopic confirmation being the case of 3C48 (Boroson & Oke 1982). However, if quasars and radio galaxies are essentially the same objects that appear to have different nuclear spectral properties because of their orientation to our line of sight and obscuration effects, then the study of host galaxies of radio galaxies is equivalent to the study of quasar hosts at the same redshifts and luminosities. The fact that a large fraction of the spatially extended rest frame UV continuum emission is *dominated* by reprocessed and scattered nuclear AGN light implies that a study of the starlight component in both radio galaxies and quasars may only be possible at near and mid infrared wavelengths.

We thank Mark Dickinson and Peter Eisenhardt for permitting us to use their image of 3C265 in this paper, and David Schlegel for his rotation curve software. We are grateful to Robert Antonucci, Sperello di Serego Alighieri, Robert Fosbury, Robert Goodrich, Todd Hurt, Buell Jannuzi, Joan Najita and Wil van Breugel for useful comments on an earlier draft of this paper. We thank Tom Bida and Wayne Wack for invaluable help during our Keck LRIS run. H.S. gratefully acknowledges NSF grant # AST-9225133. The W. M. Keck Observatory is a scientific partnership between the University of California and the California Institute of Technology, made possible by the generous gift of the W. M. Keck Foundation.

Table 1: Derived parameters for extended broad MgII emission

| Aperture Center[c] ($''$) | Extraction Width ($''$) | Flux[a] ($10^{-17}$erg s$^{-1}$ cm$^{-2}$) | FWHM ( km s$^{-1}$) | Equivalent[b] Width (Å) |
|---|---|---|---|---|
| 0.00 | 1.28 | $24.8 \pm 1.3$ | $8874 \pm 442$ | $-25.5 \pm 1.3$ |
| 0.00 | 1.71 | $30.4 \pm 1.6$ | $9126 \pm 487$ | $-26.3 \pm 1.4$ |
| 0.00 | 2.57 | $35.1 \pm 2.1$ | $9530 \pm 525$ | $-25.7 \pm 1.5$ |
| $-1.28$ | 0.86 | $2.9 \pm 0.4$ | $1839 \pm 284$ | $-17.2 \pm 2.4$ |
| $-0.86$ | 0.86 | $9.4 \pm 1.0$ | $9483 \pm 1125$ | $-29.3 \pm 3.2$ |
| $-0.43$ | 0.86 | $14.9 \pm 1.1$ | $9744 \pm 545$ | $-25.7 \pm 1.9$ |
| 0.00 | 0.86 | $18.0 \pm 1.2$ | $9474 \pm 333$ | $-25.3 \pm 1.7$ |
| $+0.43$ | 0.86 | $14.8 \pm 1.0$ | $8611 \pm 498$ | $-25.6 \pm 1.7$ |
| $+0.86$ | 0.86 | $9.1 \pm 0.8$ | $9118 \pm 1113$ | $-27.3 \pm 2.4$ |
| $+1.28$ | 0.86 | $3.5 \pm 0.6$ | $4958 \pm 1771$ | $-21.1 \pm 3.6$ |
| $+3.64$[d] | 0.86 | $7.7 \pm 1.5$ | $8857 \pm 720$ | $-43.5 \pm 8.5$ |

[a]Observed frame.

[b]Observed frame; negative equivalent widths imply emission.

[c]Measured relative to nucleus as defined by the centroid of the continuum emission along PA=150°. SE is positive.

[d]Off-nuclear knot.

# REFERENCES

Alloin, D., Arimoto, N. & Bica, E. 1989, in *Evolutionary Phenomena in Galaxies*, ed. J. E. Beckman & B. E. J. Pagel, (Cambridge: Cambridge Univ. Press), p. 411.

Antonucci, R. 1993, ARA&A, 31, 473.

Antonucci, R. J. & Miller, J. S. 1985, ApJ, 297, 621.

Baum, S. A., Heckman, T. M. & van Breugel, W. 1990, ApJS, 74, 389.

Baum, S. A., Heckman, T. M. & van Breugel, W. 1992, ApJ, 389, 208.

Boroson, T. A. & Oke, J. B. 1982, ApJ, 281, 535.

Brotherton, M. S., Wills, B. J., Steidel, C. C., & Sargent, W. L. W. 1994, ApJ, 423, 131.

Burbidge, E. M. & Burbidge, G. R. 1975, in Galaxies and the Universe, vol. 9 of Stars and Stellar Systems, ed. A. Sandage, M. Sandage, & J. Kristian, (The University of Chicago Press: Chicago) p.81.

Chambers, K.C., Miley, G.K. & van Breugel, W. 1987, Nature, 329, 604.

Cimatti, A., Alighieri, S. D., Fosbury, R. A. E., Salvati, M. & Taylor, D. 1993, MNRAS, 264, 421.

Dey, A., Cimatti, A., van Breugel, W., Antonucci, R. & Spinrad, H. 1995, ApJ, submitted.

di Serego Alighieri, S., Fosbury, R. A. E., Quinn, P. J., & Tadhunter, C. N. 1989, Nature, 341, 307.

di Serego Alighieri, S., Cimatti, A., & Fosbury, R. A. E. 1993, ApJ, 404, 584.

di Serego Alighieri, S., Cimatti, A., & Fosbury, R. A. E. 1994, ApJ, 431, 123.

Fabian, A. C. 1989, MNRAS, 238, 41P.

Henyey, L. G. & Greenstein, J. L. 1941, ApJ, 93, 70.

Jannuzi, B. T. 1994, in "Multi–Wavelngth Continuum Emission of AGN", IAU Symp. No. 159, T. J.–L. Courvoisier & A. Blecha, eds., p.470.

Jannuzi, B. T. & Elston, R. 1991, ApJ, 366, L69.

Kriss, G. A. 1994, in Proceedings of the 3rd Conference on Astrophysics Data Analysis & Software Systems, ASP Conf. Ser. v. 61, ed. D. R. Crabtree, R. J. Hanisch, & J. Barnes.

Kristian, J., Sandage, A. & Katem, B. 1978, ApJ, 219, 803.

Lilly, S. J. & Longair, M. S. 1984, MNRAS, 211, 833.

Longair, M. S., Best, P. N. & Röttgering, H. J. A. 1995, MNRAS, in press.

McCarthy, P.J. 1993, ARA&A, 31, 639

McCarthy, P.J., van Breugel, W.J.M., Spinrad, H. & Djorgovski, S. 1987, ApJ, 321, L29

Miller, J. S. & Goodrich, R. 1990, ApJ, 355, 456.

Miller, J. S., Goodrich, R. & Matthews, W. G. 1991, ApJ, 378, 47.

Oke, J. B. 1974, ApJS, 236, 27.

Oke, J. B., Cohen, J. G., Carr, M., Cromer, J., Dingizian, A., Harris, F. H., Labrecque, S., Lucino, R., Schaal, W., Epps, H., & Miller, J. 1995, PASP, 107, 375.

Osterbrock, D. E. 1989, *Astrophysics of Gaseous Nebulae and Active Galactic Nuclei*, (University Science Books: California).

Rybicki, G. B. & Lightman, A. P. 1979, *Radiative Processes in Astrophysics*, (John Wiley & Sons: New York).

Saslaw, W. C., Tyson, A. & Crane, P. 1978, ApJ, 222, 435.

Schmidt, M. 1963, Nature, 197, 1040.

Schweizer, F. 1990, in Dynamics and Interactions of Galaxies, ed. R. Wielen (Berlin: Springer-Verlag), 60.

Smith, H. E., Junkkarinen, V. T., Spinrad, H., Grueff, G. & Vigotti, M. 1979, ApJ, 231, 307.

Spinrad, H. 1986, PASP, 98, 269.

Steidel, C. C. & Sargent, W. L. W. 1991, ApJ, 382, 433.

Tadhunter, C. N., Fosbury, R. A. E., & di Serego Alighieri, S. 1988, in Proc. of the Como Conference, BL Lac Objects, ed. L. Maraschi, T. Maccacaro & M. H. Ulrich (Berlin: Springer-Verlag), 79.

Tadhunter, C. N., Fosbury, R. A. E., & Quinn, P. J. 1988, MNRAS, 240, 225.

Wilson, A. N. & Tsvetanov, Z. I. 1994, AJ, 107, 1227.

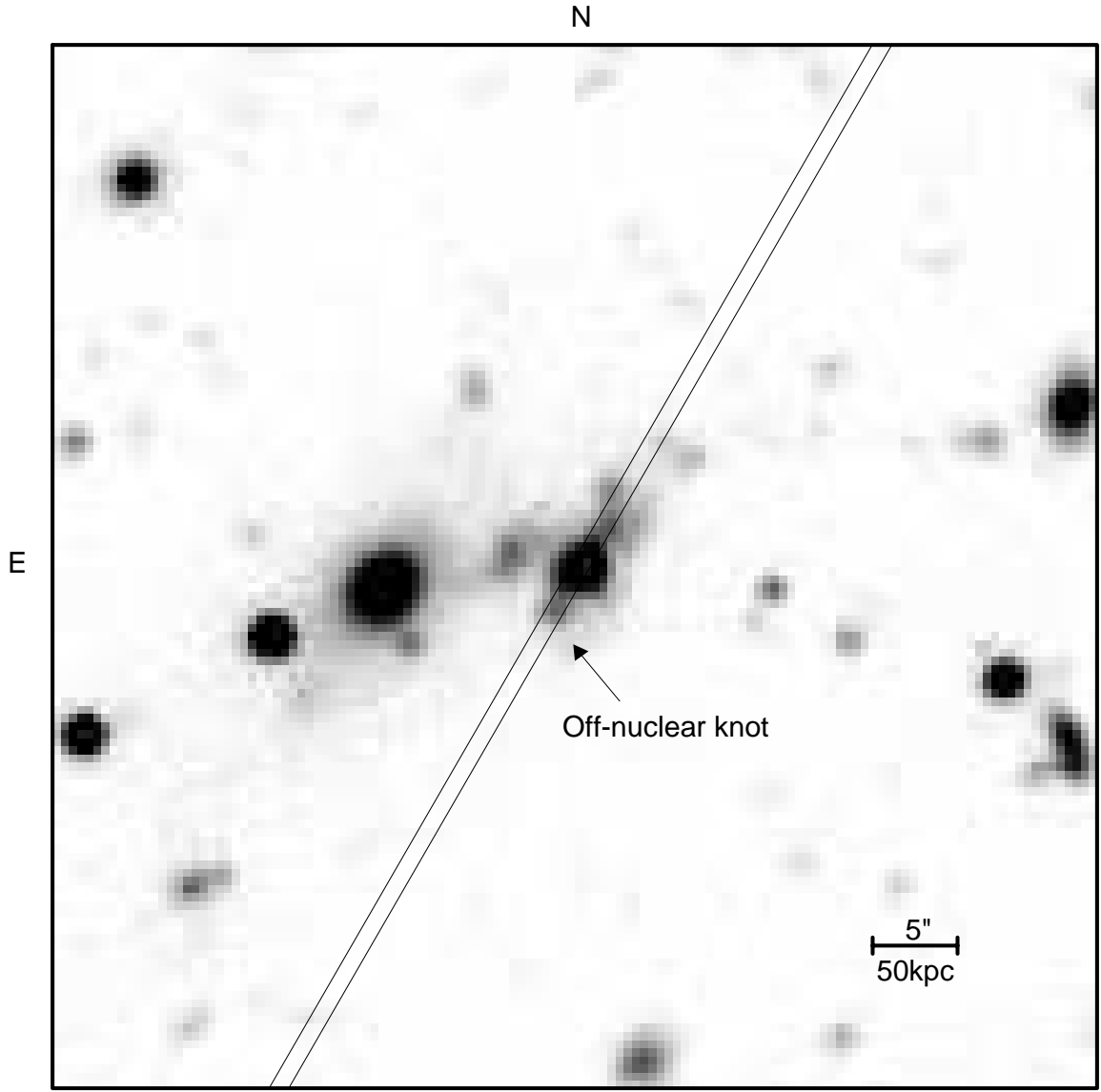

N

E

Off-nuclear knot

5"
50kpc

Fig. 1.— Broad band $R_S$ image of the 3C265 field obtained by M. Dickinson & P. Eisenhardt. The radio galaxy core position is $\alpha_{1950} = 11^h42^m52\overset{s}{.}35$, $\delta_{1950} = 31°50'26\overset{''}{.}6$. The parallel lines denote the position and orientation of the $1''$ slit used in our Keck LRIS observations. The eastern and western lobes of the radio source are oriented in PA=112° and 103° respectively. The bright galaxy to the east of 3C265 is a foreground elliptical at a redshift of z=0.392 (Smith *et al.* 1979).

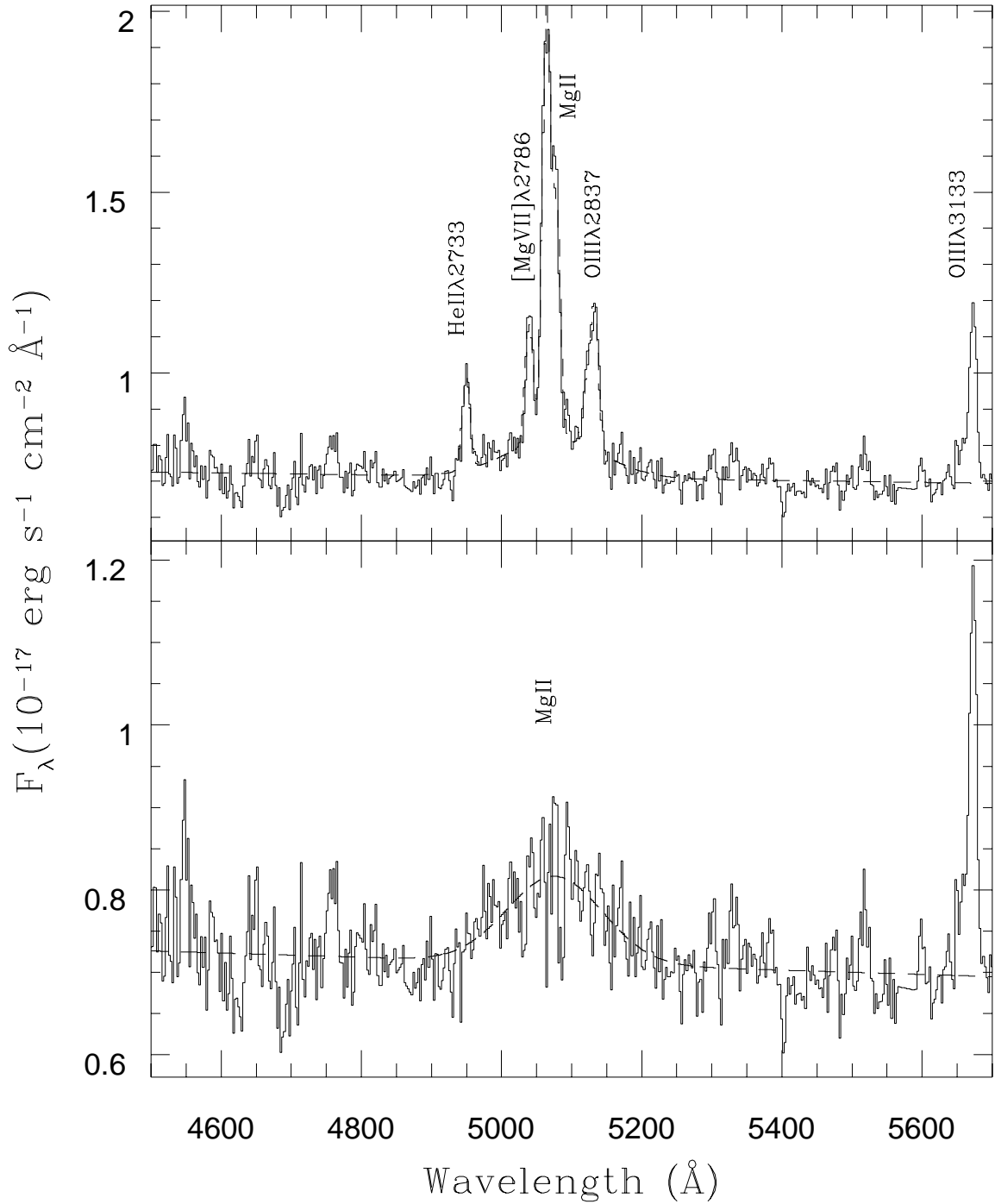

Fig. 2.— Keck LRIS spectrum of the nucleus of 3C265 showing the broad component of MgII. The extraction width is 0″86. The top panel shows the observed nuclear spectrum and a constant width gaussian model for the narrow emission lines. The bottom panel shows the broad component of the MgII line after subtraction of the narrow line model.

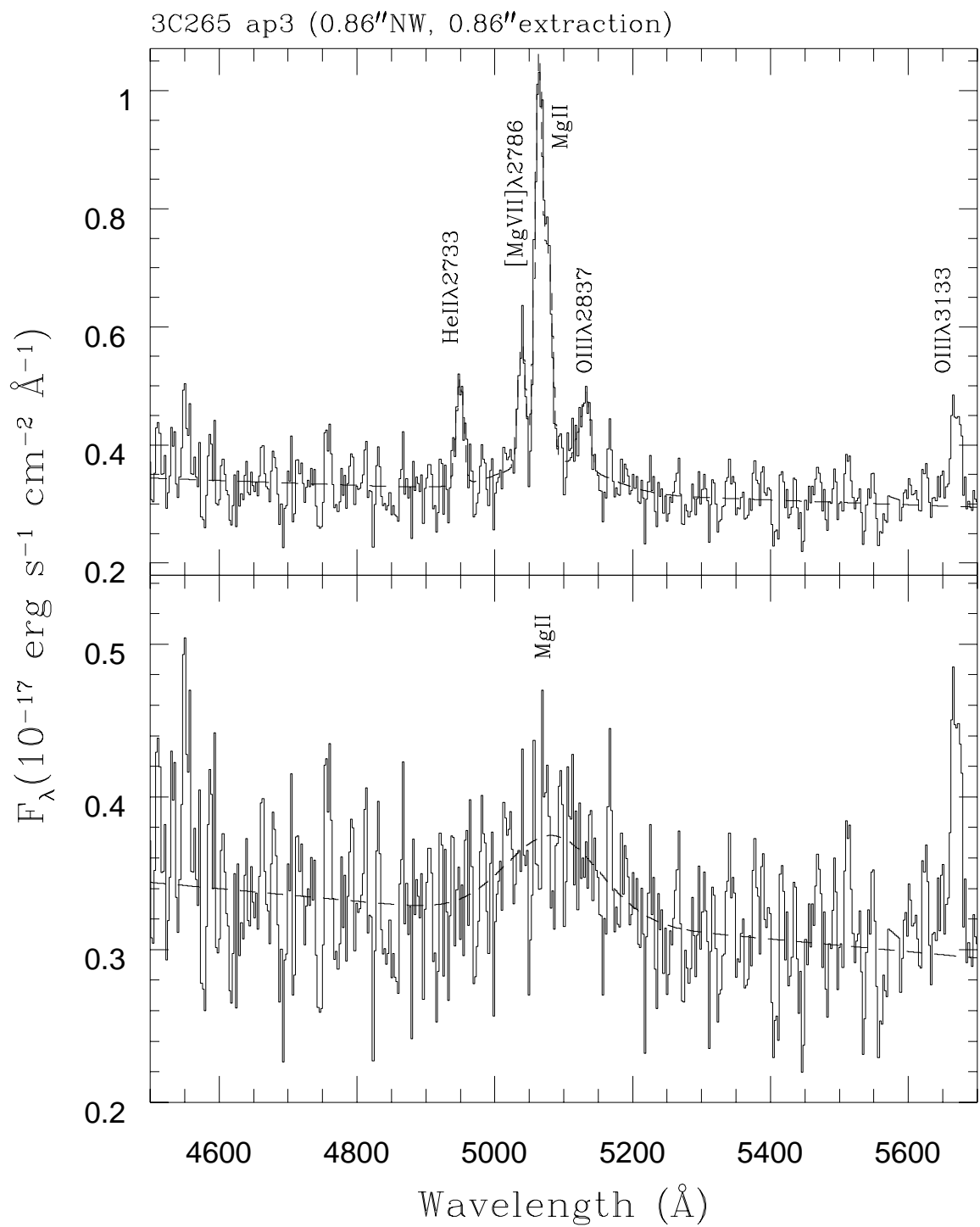

Fig. 3.— Keck LRIS spectrum of a region 0.″86 NW of the nucleus. The extraction width is 0.″86. The top panel shows the observed spectrum and a constant width gaussian model for the narrow emission lines. The bottom panel shows the broad component of the MgII line after subtraction of the narrow line model.

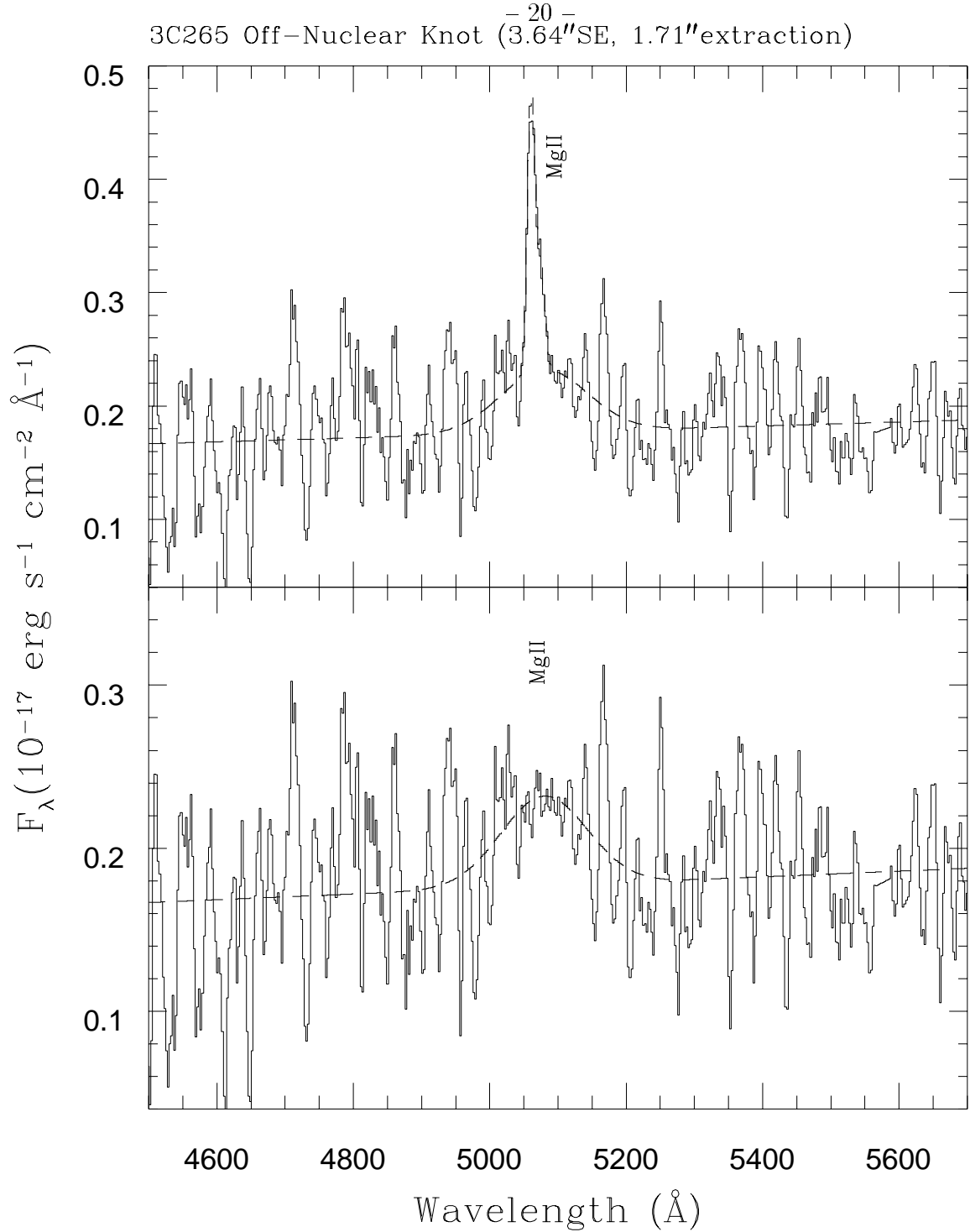

Fig. 4.— Keck LRIS spectrum of the off–nuclear knot in 3C265 showing the broad component of MgII. The extraction width of this spectrum is ≈2″ centered on a region 3″.6 away from the nucleus. The top panel shows the observed spectrum and a constant width gaussian model for the two narrow MgII emission lines. The bottom panel shows the broad component of the MgII line after subtraction of the narrow line model.

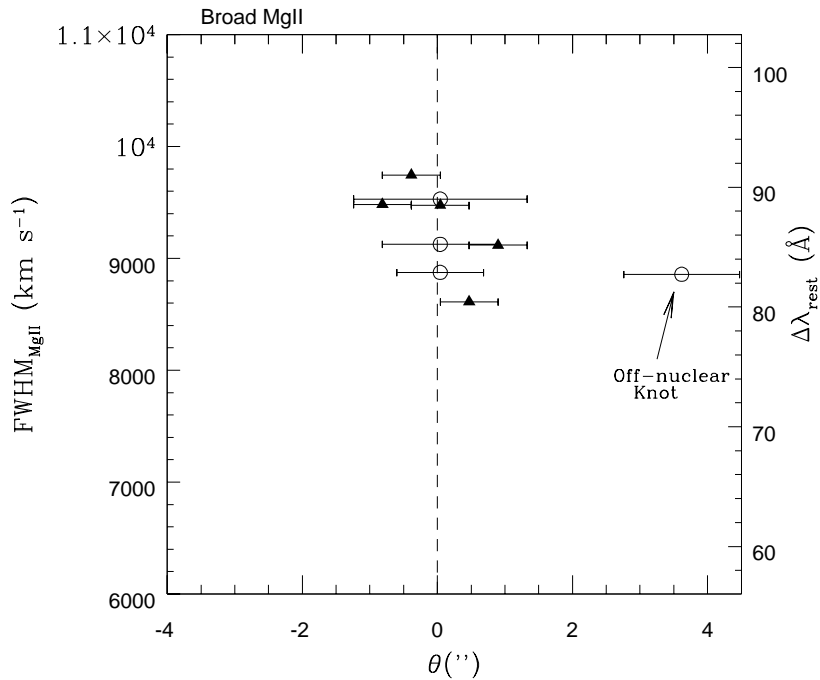

Fig. 5.— Variation of the FWHM of the broad MgII line with distance from the nucleus of 3C 265. The horizontal bars show the extraction widths of the spectra: the solid triangles denote values derived from extractions of width 0."86, and the open circles denote values derived from wider extractions of widths 1."28, 1."71 and 2."57. The formal errors are not plotted (for clarity) but listed in Table 1. Note that the broad component of the MgII line from the off-nuclear knot has roughly the same FWHM as the broad line from the nucleus, and that the width of the line ($\approx 9000\,\mathrm{km\,s^{-1}}$) is similar to quasar broad line widths.

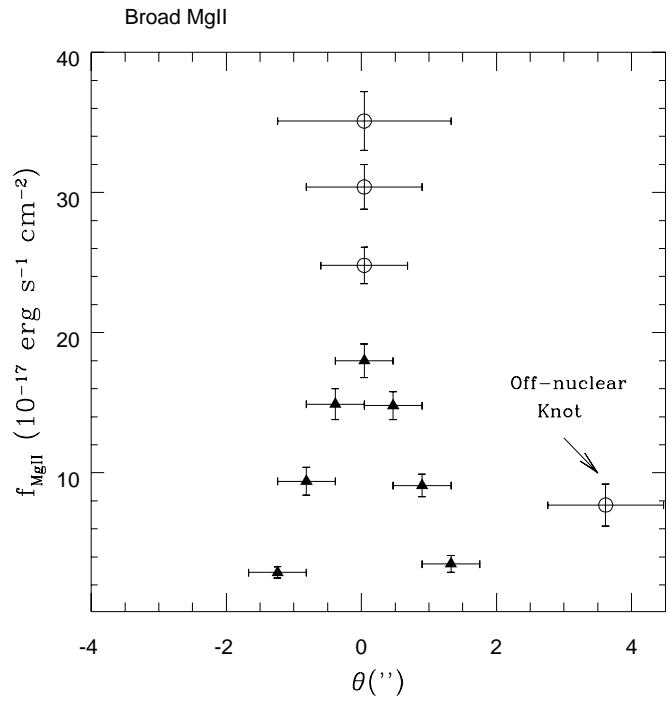

Fig. 6.— Variation of the flux in the broad MgII line as a function of distance from the nucleus of 3C265. The symbols are the same as in fig. 5. Note that the flux of the broad component increases with extraction width near the nucleus, indicating the presence of extended broad line emission in the galaxy.

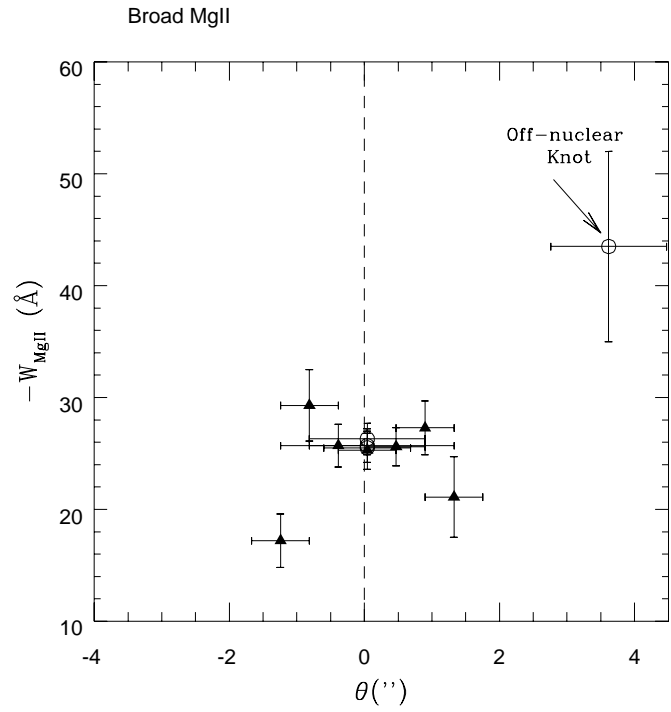

Fig. 7.— Variation of the equivalent width of the broad MgII line as a function of distance from the nucleus of 3C265. The symbols are the same as in fig. 5.

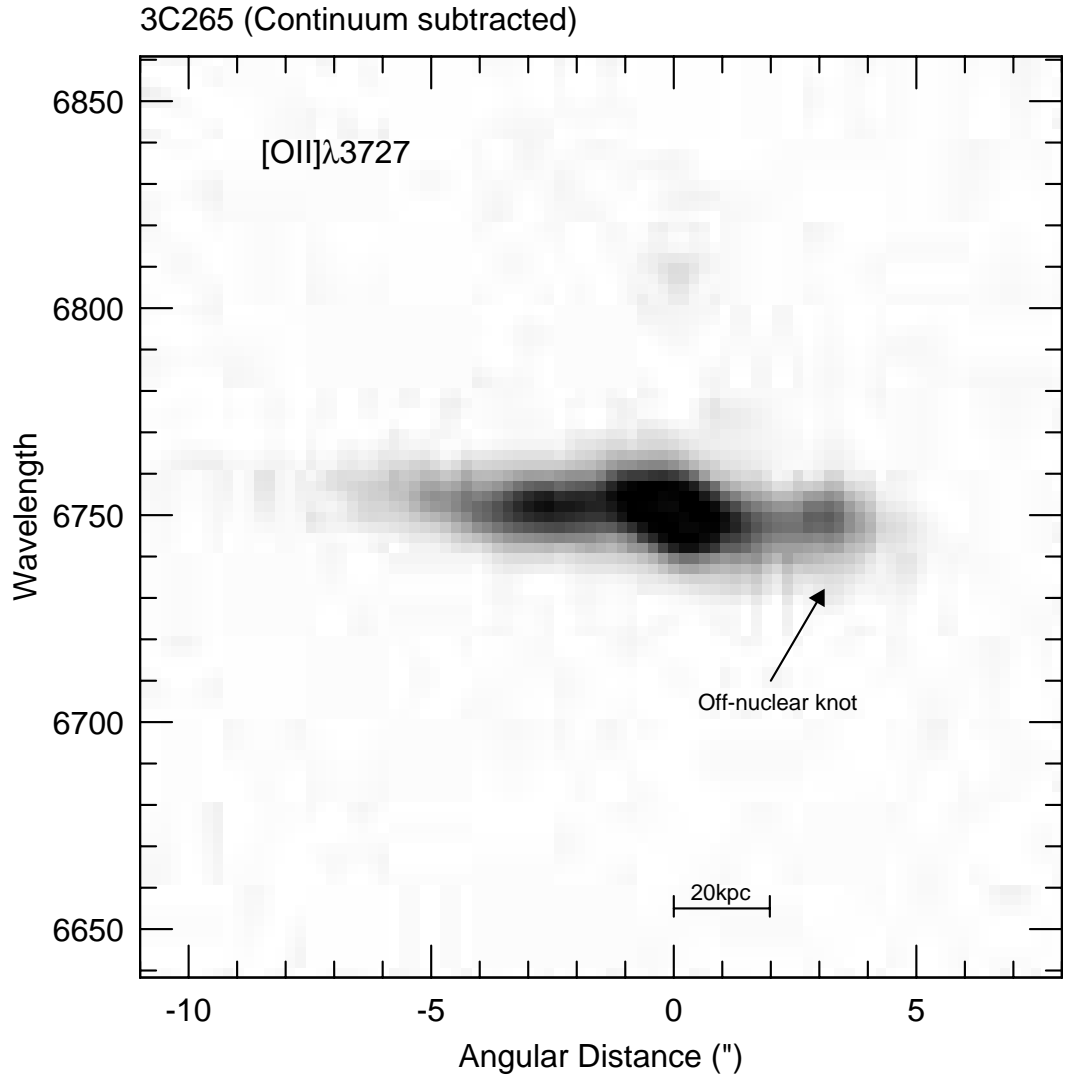

Fig. 8.— Keck LRIS two-dimensional spectrum of the [OII]λ3727 emission line in 3C265. The continuum emission of the galaxy has been subtracted from the image using a linear fit. The abscissa is the angular distance along the slit in PA=150°; θ is positive towards the SE. The origin is chosen to be the peak of the nuclear continuum emission.

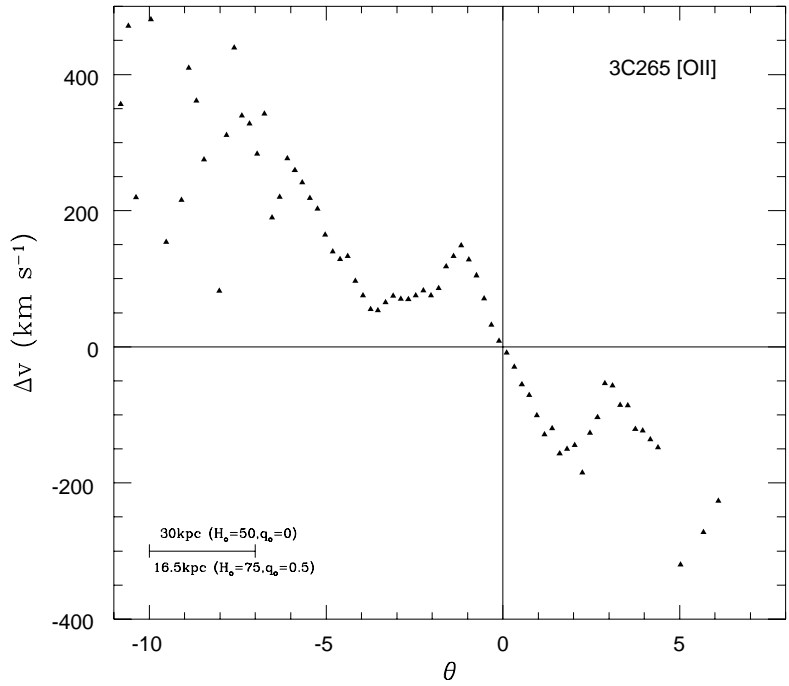

Fig. 9.— Velocity curve for 3C265 as derived from the [OII]λ3727 emission line. The abscissa is the angular distance along the slit in PA=150°; θ is positive towards the SE. The origin is defined by the location of the peak of the continuum emission from the galaxy.

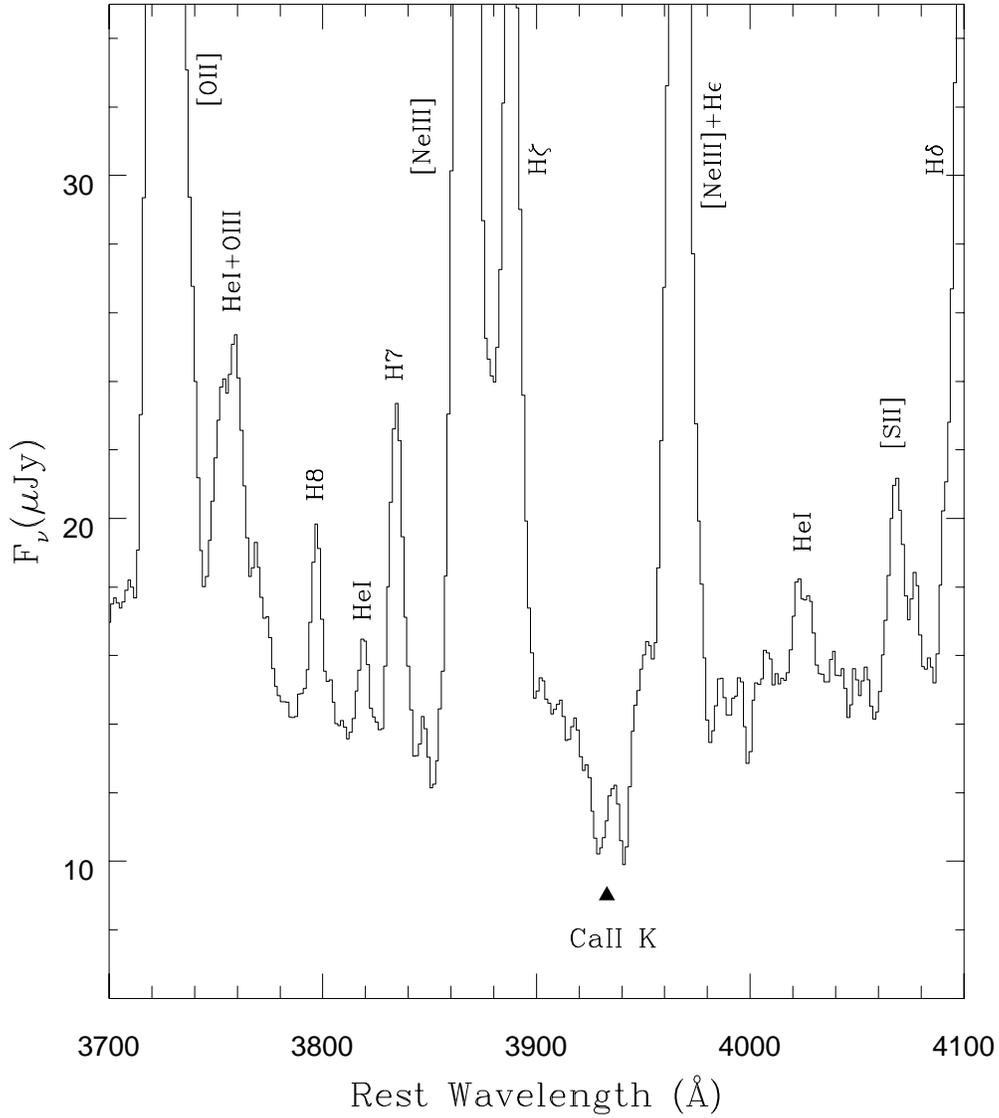

Fig. 10.— Spectrum of 3C265 in the rest frame showing the CaII K absorption line. The extraction width is ≈2″6 centered on the nucleus of the galaxy. The equivalent width of the absorption line is $> 5.5 \pm 0.5$Å and most probably a stellar feature. The CaII H line is not visible as it is swamped by the [NeIII]$\lambda$3967 and H$\epsilon$ emission lines.